\documentclass{elsart}
\def\ga{\alpha}
\def\gb{\beta}

\def\ge{\epsilon}
\def\gem{\epsilon^-}
\def\gg{\gamma}
\def\gd{\delta}

\def\gm{\mu}
\def\gn{\nu}
\def\gp{\pi}
\def\gP{\Pi}

\def\gs{\sigma}

\def\gL{\Lambda}

\def\delp{\partial_+}
\def\delm{\partial_-}

\def\part{\partial}
\def\delmu{\part_\gm}

\def\parti{\part_i}
\def\partj{\part_j}

\def\hlf{\frac{1}{2}}
\def\A0{A^{+}_0}
\def\psip{\psi_+}
\def\psin{\psi_-}
\def\psipd{\psi^{\dagger}_+}
\def\psind{\psi^{\dagger}_-}
\def\psib{\overline{\psi}}
\def\psid{\psi^{\dag}}

\def\xpl{x^{+}}
\def\xmin{x^{-}}
\def\ymin{y^{-}}
\newcommand{\nc}{\newcommand}
\nc{\intgl}{\int\limits_{-L}^{+L}\!{{dx^-}\over\!2}}
\nc{\intgly}{\int\limits_{-L}^{+L}\!{{dy^-}\over\!2}}
\nc{\zmint}{\int\limits_{-L}^{+L}\!{{dx^-}\over{\!2L}}}
\def\xperp{x^\perp}

\def\pperp{p^\perp}

\nc{\uli}{\underline}
\def\ulix{\uli{x}}

\def\ulip{\uli{p}}

\def\ulige{\uli{\ge}}
\def\beq{\begin{equation}}
\def\eeq{\end{equation}}
\def\bea{\begin{eqnarray}}
\def\eea{\end{eqnarray}}
\usepackage{graphics}
\begin{document}
\begin{frontmatter}
\hyphenation{}
\title{Light-Front Aspects of Chiral Symmetry} 
\author{L$\!\!$'ubom\'{\i}r Martinovi\v c$^{a,c}$ and James P. Vary$^{b,c}$ } \\
\address{$^a$Institute of Physics, D\'ubravsk\'a cesta 9, 842 28 
Bratislava, Slovakia\\
$^b$Department of Physics and Astronomy, Iowa State University \\
$^c$International Institute of Theoretical and Applied Physics, \\
Iowa State University, Ames, Iowa 50011, USA}
\date{}

\begin{abstract}
Spontaneous chiral symmetry breaking and axial anomaly are studied in the 
light-front formulation. The existence of multiple vacua and a Nambu-Goldstone 
boson, both related to dynamical fermion zero modes, are demonstrated within a 
simple sigma model with fermions. The Weyl gauge formulation and a 
consistent gauge invariant point-splitting regularization, which includes 
the light front time, are crucial for obtaining the anomaly in the massive 
Schwinger model and QED(3+1). 
\end{abstract}
\end{frontmatter}

\section{Introduction}

Relativistic field theory quantized at the equal light-front time $\xpl$ 
$(x^\pm = z\pm t)$ \cite{LKS} is both promising and puzzling. Its 
primary advantage -- a kinematical definition of the physical vacuum 
(modulo zero modes), seems to be in conflict with rich physical contents of the 
vacuum in the usual space-like quantization.  
Also, another fundamental property of theories with fermions, namely   
chiral symmetry \cite{PRC,Burk}, has not been fully understood 
in the light-front (LF) framework. For example, non-zero   
chiral condensates of the Schwinger model have not been obtained within  
the genuine LF quantum field theory, whose structure differs in many important 
aspects from the usual space-like theory. 

In the present contribution, we will study two aspects of 
chiral symmetry on the light front: spontaneous symmetry breaking and the 
axial anomaly. First, we will sketch the relation between the dynamical 
fermion zero modes and the existence of the Goldsone boson in a simple 
$O(2)$-symmetric sigma model with fermions. The axial anomaly will be  
analyzed in two and four-dimensional quantum 
electrodynamics in the Weyl-gauge formulation, using a consistent gauge 
invariant regularization by splitting in LF space {\it and time} variables. 

\section{Goldstone theorem on the light front}

Spontaneous symmetry breaking is usually associated with non-invariance 
of the vacuum state of a given Hamiltonian. If there are also non-invariant 
(composite) operators $A$ with the property $\langle 0 \vert \gd A \vert 0 
\rangle \neq 0$, existence of a massless state(s) with the quantum 
numbers of the conserved current(s), corresponding to the symmetry group, 
can be proved. Due to the simplicity of the physical LF vacuum in the sector 
of normal modes, this picture can be realized on the LF explicitly in the Fock 
representation using a finite-volume formulation. The simplest example is a 
model without gauge symmetry, namely 
the $O(2)$-symmetric sigma model with periodic massless fermions \cite{LJ}.  
Unitary operators $V(\gb)=\exp\big(-i\gb Q_5\big)$, implementing the axial 
symmetry of the model, contain a part composed of dynamical fermion zero 
modes $b_0,d_0$, and generate a set of degenerate vacua 
$ \vert \gb \rangle = \exp\Big[-i\gb\sum_{s}2s\big[b^\dagger_0(s)d^\dagger_0
(-s) + h.c.\big]\Big]\vert 0 \rangle$, where $s$ is the LF helicity.
There are two non-invariant operators $A$: $\psib\psi = 
\psipd\gg^0\psin + h.c.$ and $ \psib\gg^5 \psi=\psipd\gg^0\gg^5\psin + h.c.$,  
$A \rightarrow 
V(\gb)AV^\dagger(\gb) \neq A$, and the conserved axial vector current 
$j^\gm_5 = \psid\gg^0\gg^\gm\gg^5\psi~ +$ bosonic part ($\gm=+,-,1,2$, see 
Sec.4 for our notation). 
Thus, the existence of a massless state $\vert G \rangle = Q_5\vert 0 \rangle = 
\sum_{s}2sb^\dagger_0(s)d^\dagger_0(-s)\vert 0 \rangle$ can be derived in 
a usual way (see \cite{IZ}, e.g.). 
   
\section{Axial anomaly in the massive Schwinger model}

The LF Hamiltonian of the massive Schwinger model in the gauge $A^-=0$ 
\beq
P^- = \int d\xmin\left[\gP_{A^+}^2 + m\left(\psipd\gg^0\psin + \psind\gg^0\psip
\right)\right]
\label{Wham}
\eeq
is expressed in terms of the conjugate momentum $\gP_{A^+}=\delp A^+$ of the 
$A^+$ gauge field component and dynamical (+) and dependent (--) fermi fields 
$\psi_{\pm}$. We work in the continuum formulation and all fields are taken 
to be antiperiodic in $\xmin$ \cite{Steinh}. The dynamical component $\psip$ 
obeys the equation $ 2i\delp\psip = m\gg^0\psin$ and is expanded at $\xpl=0$ as 
\bea
\psip(\xmin) = \int\limits_0^{\infty}{dp^+ \over {4\gp \sqrt{p^+}}}
u \left(b(p^+)e^{-{i \over 2}p^+\xmin} + d^{\dagger}(p^+)e^{{i \over 2}p^+\xmin}
\right),\;\;u^\dagger=(0~1),\nonumber \\
\{b(p^+),b^{\dagger}(p^{\prime +})\} = \{d(p^+),d^{\dagger}(p^{\prime +})\} = 
4\gp p^+ \gd(p^+ - p^{\prime +}).
\label{contexp}
\eea 
The dependent component $\psin$ in (\ref{Wham}) satisfies the constraint  
($x=(\xpl,\xmin)$)
\beq
2i\delm\psin(x) = m\gg^0\psip(x) + e \psin A^+(x),
\label{psineq}
\eeq 
which is inverted by the antiperiodic Green's function 
$G_a (\xmin-\ymin;A^+)$$=\hlf \ge_a(\xmin\!-\ymin)\exp{\Big[-{{ie}\over 2} 
\int\limits_{\ymin}^{\xmin}\!\!dz^-A^+(\xpl,z^-)\Big]}$, 
where $\ge_a(\xmin)$ is the sign function:   
\beq  
\psin(\xpl,\xmin)={m \over {2i}}\gg^0\int dy^-G_a(\xmin-\ymin;A^+)
\psip(\xpl,\ymin). 
\label{psinv}
\eeq 

For a successful calculation of the axial anomaly, it is crucial to work with 
a consistent definition of the fermion currents. On physical grounds, the 
vector current has to be odd under C-parity \cite{BD}. This can be achieved by 
taking an antisymmetrized product of fermi fields which is equivalent 
to normal ordering and thus removes the short-distance singularity in the 
product even if a gauge invariant splitting of the arguments of the field 
operators \cite{Schw} is performed. 
On the other hand, there is no C-parity restriction in the case of the axial 
current $j^\gm_5 = \psid\gg^0\gg^\gm\gg^5\psi$. Its regularized and gauge 
invariant definition is  
\beq
j^{\pm}_5(x) = \pm 2 \psi^\dagger_{\pm}(\xpl\!+{\gd \over 2},\xmin\! + 
{\gem \over 2}) \psi_{\pm}(\xpl\!-{\gd \over 2},\xmin \!-{\gem\over 2})  
e^{-{ie\over 2}\!\!\!\int \limits_ {x-{\ge / 2}}^{x+ 
{\ge / 2}}\!\!\!d\ymin A^+(\ymin)}.
\label{axplus}
\eeq
The regulators $\ge=(\gem,\gd)$ are set to zero at the end of calculations.  
Now, using the equation of motion for an appropriate choice of split  
arguments, we get 
\bea 
\delp j^+_5(x) = im\Big[\psind(x)\gg^0 \psip(x) - h.c.\Big] 
e^{-{ie \over 2}\gem A^+(x)}
- {ie\over 2} \gem \delp A^+(x) j^+_5(x).
\label{delpj5}
\eea 
Naively, it looks like the second term above vanishes. Indeed, using the 
anticommutator (\ref{contexp}), one finds 
$j^+_5= :j^+_5: - {e \over {2\gp}} A^+$ which, however,  
is not gauge invariant (the second term gets shifted by the residual 
time-independent gauge transformation). The only way out appears to be the use 
of a gauge-invariant version of the anticommutator :       
\beq
\{\psip(\xmin),\psipd(\ymin)\} = \hlf \gL_+ \exp{\Big({ie \over 2}\int\limits_
{\xmin}^ {\ymin}dz^- A^+(z^-)\Big)} \gd_{a}(\xmin-\ymin).
\label{gACR}
\eeq
and the corresponding change of the anticommutators (\ref{contexp}). With this 
modification, which is analogous to the Schwinger's current definition,  
the vector current $j^{\pm}(x) = 2: \psi^{\dagger}_{\pm}(x)\psi(x)_{\pm} :$  
is conserved, $j^+_5(x) = :j^+_5(x): + {1 \over{i\gp\gem}}$ and 
\beq
\delp j^+_5(x) = im\Big[\psind(x)\gg^0\psip(x)  
- \psipd(x)\gg^0\psin(x)\Big]
- {e\over 2\gp} \delp A^+(x).
\label{plusano}
\eeq 
For the case of the minus (``bad'') component of the vector current, we use  
the constraint (\ref{psineq}) for the appropriate 
choice of arguments and expand in $\gem,\gd$: 
\bea
\delm j^-_5(x) =  im\Big[\psind(x)\gg^0\psip(x) - h.c.\Big] 
e^{-{ie\gem \over 2} A^+(x)} + {ie\over 2} \gd \delp A^+(x)j^-_5(x).
\label{del5}
\eea 
The second term can give a non-zero contribution if the $j^-_5$ current  
has a singularity $\gd^{-1}$. To investigate this 
point, we insert $\psin$ (\ref{psinv}) into the definition of $j^-_5$. 
The time dependence of the field $\psipd(\xpl + {\gd \over 2},x^-)$ is  
assumed to be of the free-field form $\exp{\big(\pm{i \over 2}\hat{p}^-
\gd\big)}$, where $\hat{p}^-= {m^2 \over p^+}$. Normal-ordering $j^-_5$ 
and using the gauge invariant anticommutator (\ref {gACR}), we find for 
the contraction part  

\beq
<j^-_5(x)> = -m^2\int\limits_{0}^{\infty}{dp^+ \over p^{+2}}e^{-{i \over 2}
p^+\xmin - {i \over 2}{m^2 \over p^+}\gd} = -{1 \over {i\gp\gd}}, 
\label{timesing}
\eeq
where the expansion of the modified Bessel function 
$K_1(m\sqrt{\gem\gd})$ for a small argument has been used. 
Thus, the mass dependence completely cancels \cite{Berg} and the second term in 
Eq.(\ref{del5}) gives a finite contribution. In this way, the divergence of 
the axial vector current in the massive LF Schwinger model in the Weyl gauge 
is 
\beq
\partial_\gm j^\gm_5 = \delp j^+_5+\delm j^-_5 = 2im\Big[\psind\gg^0
\psip - \psipd\gg^0\psin\Big] - {e \over \gp}\delp A^+. 
\label{anomaly}
\eeq
This result includes also the massless case. The anomaly (\ref{anomaly}) can 
be written in the usual covariant form ${e \over {2 \gp}}\ge^{\gm\gn}
F_{\gm\gn}$ $(\ge^{+-} = -2, A^-=0)$.   

\section{Anomaly of the LF QED(3+1)}

The calculation of the axial anomaly in the four-dimensional $U(1)$ theory 
proceeds essentially as in the two-dimensional model. In particular, one has 
to use a four-dimensional analogue of the ``gauge-corrected'' 
anticommutator (\ref{gACR}). The obvious difference is the presence of the 
two perpendicular gauge field components $A^i(x)$ (interpreted as a 
classical background field \cite{Jackiw}) and the spin degree of freedom of 
the fermi field.
The field equations are 
\bea
2i\delp\psip &=& \big(m\gg^0 - i \ga^i \parti \big)\psin - e \ga^i A^i\psin,
\\
2i\delm\psin &=& \big(m\gg^0 - i \ga^i \parti \big)\psip - e \ga^i A^i\psip + 
e A^+\psin .
\label{EM}
\eea  
The solution of the constraint (\ref{EM}) is again given by the Green's 
function $G_a(\xmin-\ymin;A^+)$: 
$\psin(x) = \int {dy^- \over 4i} G_a(\xmin-\ymin;A^+)
\Big[m\gg^0 - i\ga^i\parti -e\ga^i A^i(y) \Big]\psip(y)$.
Our notation is $\ulix=(\xmin,x^\perp), y=(\xpl,\ymin,x^\perp), 
x^\perp=x^i, \ga^i=\gg^0\gg^i, i=1,2$. 
\hspace{2mm}
Let us calculate the divergence of the axial vector current \\ 
$j^\mu_5(x) = \psid(\xpl\!+{\gd \over 2},\ulix + {\ulige \over 2})\gg^0\gg^\gm
\gg^5\psi(\xpl\!-{\gd \over 2},\ulix - {\ulige \over 2})
\exp{\big(\!-ie\!\!\! \int\limits^{x + {\ge / 2}}_{x + {\ge / 2}}\!\!\!\!
dy_\ga A^\ga(y) \big)}$: 
\hspace{-2mm}
\bea
&&\delmu j_5^\gm(x) = 2im:\Big[\psind(x)\gg^0\gg^5 \psip(x)  
- h.c. \Big]: + imC_m - im\overline{C}_m + C + \overline{C} \nonumber \\
&&+  C^iA^i(x+ {\ge \over 2}) - \overline{C}^iA^i(x-{\ge \over 2}) 
- ie\ge^\gm \Big[\parti A_\gm\langle j^i_5(x)\rangle +  
\delm A_\gm(x) \langle j^-_5 \rangle\Big] \nonumber \\ 
&&+{ie \over 2} <j^->\Big[A^+(x + {\ge \over 2}) - A^+
(x - {\ge \over 2})\Big]  
\eea
Here, the field equations and normal ordering have been used to isolate the 
contractions  
$C_m = \Big< \psipd(x + {\ge \over 2})\gg^0 \gg^5 \psin(x-
{\ge \over 2}) \Big>,\; 
C = \Big<\psipd(x + {\ge \over 2})\ga^i\gg^5 \parti 
\psin(x - {\ge \over 2}) \Big >,
C^i = -ie\Big<\psipd(x + {\ge \over 2})\ga^i\gg^5\psin(x - 
{\ge \over 2})\Big>$, with $\overline{C}=C^*(-\ge)$, etc. 
They can be evaluated by simple spinor identities. $C_m$  
vanishes and the remaining contractions lead, due to the splitting in $\xpl$,   
to Bessel-type integrals, as for example 
\beq
\int {d^3\ulip \over{k^+ - p^+}}e^{-{i \over 2}p^+\gem +i\pperp\xperp - 
{i \over 2}{(m^2 + p_\perp^2) \over p^+}\gd} = -{i \over {8\gp}}{k^+ 
\over \gd}e^{-{i \over 2}{k^+(\gem + {\ge_\perp^2 \over \gd})}+{i \over 2}{m^2 \over k^+}\gd}.
\eeq
Expanding the gauge potentials in $\gd$, one finds finite contributions to the 
divergence of the axial vector current. From the $C^iA^i$ terms, e.g., we get 
$e^2/(2\gp^2)\ge^{ij}\delm A^j\delp A^i$ $(\ge^{ij}=-\ge^{ji})$ which is a 
part of the complete result. Indeed, the full abelian anomaly in the covariant 
form is \cite{Jackiw}
\beq
\delmu j^\gm_5(x) = -{e^2\over {16\gp^2}}\ge^{\gm\gn\rho\gs}
F_{\gm\gn}F_{\rho\gs}.
\label{covanom}
\eeq
Its LF form in the $A^- = 0$ gauge is $e^2
/(2\gp^2)\ge^{ij}\big[\delp A^+\parti A^j + \delp A^i \delm A^j + 
\hlf \delp A^i 
\partj A^+ \big]$. The second term coincides with our partial result above. 
The complete treatment with all technical details will be given elsewhere.

\end{document}